\documentclass[%
 aip,
 amsmath,amssymb,
 reprint,%
]{revtex4-1}

\usepackage{graphicx}% Include figure files
\usepackage{dcolumn}% Align table columns on decimal point
\usepackage{bm}% bold math
\usepackage[utf8]{inputenc}
\usepackage[T1]{fontenc}
\usepackage{mathptmx}
\usepackage{overpic} 
\usepackage{tikz} 

\begin{document}

\preprint{AIP/123-QED}
\title[Network synchronization]{Dense networks that do not synchronize and sparse ones that do}
%\title[Circulant networks]{Circulant networks of identical Kuramoto oscillators: Seeking dense networks that do not globally synchronize and sparse ones that do}
% Force line breaks with \\
\author{Alex Townsend}
\altaffiliation{Department of Mathematics, Cornell University, Ithaca, NY 14853}%
\author{Michael Stillman}
\altaffiliation{Department of Mathematics, Cornell University, Ithaca, NY 14853}%
\author{Steven H. Strogatz}
\altaffiliation{Department of Mathematics, Cornell University, Ithaca, NY 14853}%

%\author{A. Author}
%\author{B. Author}%
% \email{Second.Author@institution.edu.}
%\affiliation{ 
%Authors' institution and/or %address%\\This line break forced with \textbackslash\textbackslash
%}%

% \homepage{http://www.Second.institution.edu/~Charlie.Author.}
%\affiliation{%
%Second institution and/or address%\\This line break forced% with \\
%}%

\date{\today}% It is always \today, today,

\begin{abstract}
For any network of identical Kuramoto oscillators with identical positive coupling, there is a critical connectivity above which the system is guaranteed to converge to the in-phase synchronous state, for almost all initial conditions. But the precise value of this critical connectivity remains unknown. In 2018, Ling, Xu, and Bandeira proved that if each oscillator is coupled to at least 79.29 percent of all the others, global synchrony is ensured. In 2019, Lu and Steinerberger improved this bound to 78.89 percent. Here, by focusing on circulant networks, we find clues that the critical connectivity may be exactly 75 percent. Our methods yield a slight improvement on the best known lower bound on the critical connectivity, from $68.18\%$ to $68.28\%$. We also consider the opposite end of the connectivity spectrum, where the networks are sparse rather than dense. In this regime, we ask how few edges one needs to add to a ring of $n$ oscillators to turn it into a globally synchronizing network. We prove a partial result: all the twisted states in a ring of size $n=2^m$ can be destabilized by adding just $\mathcal{O}(n \log_2 n)$ edges. To finish the proof, one also needs to rule out all other candidate attractors. We have done this for $n=8$ with computational algebraic geometry, but the problem remains open for larger $n$. Thus, even for systems as simple as Kuramoto oscillators, much remains to be learned about dense networks that do not globally synchronize and sparse ones that do.    
\end{abstract}

\maketitle

\begin{quotation}
The dynamics of a system of oscillators can be profoundly affected by its wiring diagram. For example, consider a network of $n$ identical Kuramoto oscillators, all coupled to their neighbors with unit strength. The network is assumed to be connected but otherwise has arbitrary topology. Under what conditions can we be sure the oscillators will settle into a state of in-phase synchrony instead of another mode of organization? A few things are known: if the oscillators are densely coupled all-to-all, the system globally synchronizes (meaning it asymptotically approaches the in-phase state from almost all initial conditions). On the other hand, if the oscillators are arranged in a ring and sparsely coupled only to their nearest neighbors, the system can display other patterns besides in-phase synchrony; these patterns take the form of twisted states in which the oscillators' phases differ by a constant amount from one to the next around the ring. To clarify how the topology of a network affects its propensity to synchronize, we analyze circulant networks of identical Kuramoto oscillators. For these highly symmetric networks, the linear stability of the twisted states can be solved completely. The results lead us to conjecture that global synchronization occurs in any network of identical Kuramoto oscillators in which each oscillator is connected to more than 75 percent of the others. At the sparse end of the connectivity spectrum, we provide evidence that a ring of $n$ oscillators can be turned into a globally synchronizing network by adding as few as $\mathcal{O}(n \log_2 n)$ edges in the right places. These results could be potentially useful in the design of distributed systems of clocks, sensors, low-power radios, or other applications where many oscillators need to keep themselves in sync without any cues from a central timekeeper.
\end{quotation}

\section{\label{sec:intro}Introduction}
Coupled nonlinear oscillators often fall into sync spontaneously. Examples range from flashing fireflies and neural populations to arrays of Josephson junctions and nanoelectromechanical oscillators~\cite{winfree1967biological,peskin1975mathematical, kuramoto1984chemical, mirollo1990synchronization, watanabe1994constants, pikovsky2003synchronization, matheny2019exotic}. On the theoretical side, many researchers have explored how the tendency to synchronize is affected by network structure~\cite{jadbabaie2004stability, wiley2006size,mallada2010synchronization, taylor2012there, dorfler2014synchronization,pecora2014cluster, pikovsky2015dynamics, canale2015exotic, Mehta15, rodrigues2016kuramoto,abrams2016introduction,deville2016phase, sokolov2018sync, ling2018landscape,lu2019synchronization}. This is the topic of the present paper.

We say that a system of oscillators \emph{globally synchronizes} if it converges to a state with all the oscillators in phase, starting from all initial conditions except a set of measure zero. Intuitively, one expects that dense networks should favor global synchronization, whereas sparse networks might support waves and other patterns besides synchrony. In 2012, Taylor~\cite{taylor2012there} explored these issues using a system of identical Kuramoto oscillators. He proved that the system globally synchronizes if each oscillator is coupled to at least 93.95\% of the others. This result raised a natural question: What is the critical connectivity for systems of identical Kuramoto oscillators? To make this notion precise, define the \emph{connectivity} $\mu$ of a network of size $n$ as the minimum degree of the nodes in the network, divided by $n-1$, the total number of other nodes. Then define the \emph{critical connectivity} $\mu_c$ as the smallest value of $\mu$ such that any network of $n$ oscillators is globally synchronizing if $\mu\geq\mu_c$; otherwise, for any $\mu < \mu_c$, at least one network exists with some other attractor besides the in-phase state. In these terms, Taylor's result is $\mu_c \le 0.9395$. 

The striking thing about Taylor's result is that the structure of the network could be arbitrary in all other respects (random, regular, or in between). Recently, Ling, Xu, and Bandeira~\cite{ling2018landscape} refined Taylor's argument to show that 79.29\% connectivity ensures global synchronization, and Lu and Steinerberger~\cite{lu2019synchronization} lowered the bound still further to 78.89\%. On the other hand, competing attractors called twisted states~\cite{wiley2006size,canale2015exotic} can coexist with the synchronous state for certain networks whose connectivity is less than 68.18\%. Thus  $0.6818\leq \mu_c \leq 0.7889$. 

\begin{figure} 
\tikzstyle{network}=[circle, draw, fill=black!50,
                        inner sep=0pt, minimum width=4pt]
\begin{minipage}{.155\textwidth}
\includegraphics[]{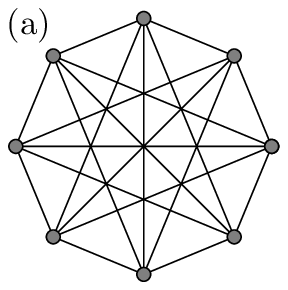}
\end{minipage}
\begin{minipage}{.155\textwidth}
\includegraphics[]{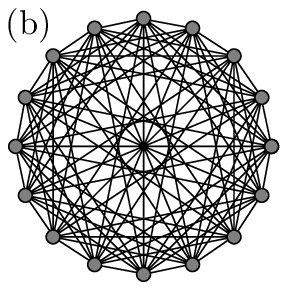}
\end{minipage}
\begin{minipage}{.155\textwidth}
\includegraphics[]{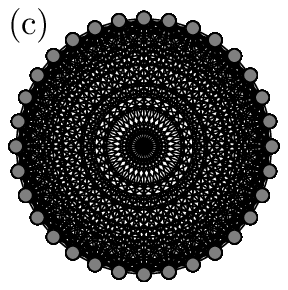}
\end{minipage}

\begin{minipage}{.155\textwidth}
\includegraphics[]{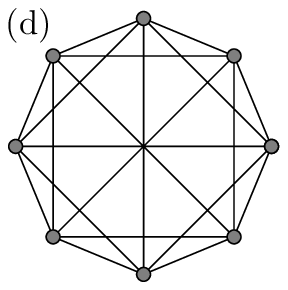}
\end{minipage}
\begin{minipage}{.155\textwidth}
\includegraphics[]{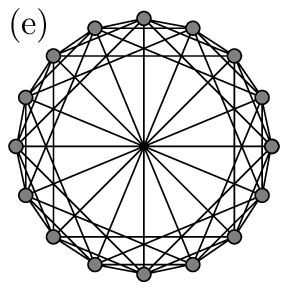}
\end{minipage}
\begin{minipage}{.155\textwidth}
\includegraphics[]{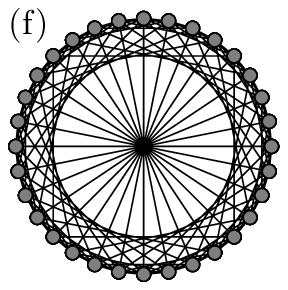}
\end{minipage}
\caption{Dense circulant networks and sparse ones of size $n = 8$, $16$, and $32$. Panels (a)-(c) show networks with a twisted state whose associated Jacobian matrices have all eigenvalues $\leq 0$. These networks are tantalizingly close to not being globally synchronizing. We wonder if they could be tweaked to be truly so. Panels (d)-(f) show sparse circulant networks that have no stable twisted states. These are promising candidates for the sparsest circulant networks that globally synchronize, but we cannot say that for certain, because we have not ruled out the possible existence of other attractors.} 
\label{networkExamples} 
\end{figure}

In this paper, we use symmetric structures called circulant networks to sharpen the understanding of $\mu_c$. We construct a family of networks (Fig.~\ref{networkExamples}, top row) whose connectivity approaches 75\%, and which lie on the razor's edge of being able to avoid global synchrony: along with the synchronous state, they have a competing, nearly stable, twisted state. In each case, that twisted state has all negative eigenvalues, apart from four zero eigenvalues which render linear analysis insufficient. Simulations show that, sadly, this twisted state is weakly \emph{unstable}. We had hoped it would be stable, because that would have implied $\mu_c \ge 0.75$. At least we can say that any future attempt to show $\mu_c < 0.75$ by linear analysis (in the style of Taylor~\cite{taylor2012there} and his successors~\cite{ling2018landscape, lu2019synchronization}) must now contend with this sequence of networks and its zero eigenvalues. Its near-miss quality leads us to conjecture that $\mu_c = 0.75$. There is also a surprise at the opposite end of the connectivity spectrum (Fig.~\ref{networkExamples}, bottom row). Merely connecting each oscillator to a logarithmically small number of neighbors suffices to destabilize all the twisted states of a ring, thereby converting it (we conjecture) into a globally synchronizing network. 

To find these example networks, we consider a homogeneous Kuramoto model in which each oscillator has the same  natural frequency (which can be set to zero without loss of generality, by going into a suitable rotating frame). The governing equations are 
\begin{equation} 
\frac{d\theta_j}{dt} = \sum_{k=0}^{n-1} A_{jk} \sin\!\left(\theta_k - \theta_j\right), \quad 0\leq j\leq n-1. 
\label{eq:dynamical}
\end{equation} 
Here the entries of the adjacency matrix $A_{jk} = A_{kj} = 1$ if oscillator $j$ is coupled to oscillator $k$; thus, all interactions are symmetric, equally attractive, and  normalized to unit strength. Since the adjacency matrix $A$ is symmetric,~\eqref{eq:dynamical} is a gradient system~\cite{wiley2006size,jadbabaie2004stability}. Thus all the attractors are equilibrium points, so to analyze the long-term global dynamics of~\eqref{eq:dynamical}, it suffices to analyze the local stability of equilibria. 

\section{Circulant networks} 
Circulant networks turn out to be a rich source of graphs for our purposes. A circulant network is a graph whose adjacency matrix $A$ has constant diagonals, such that  each row is a circularly shifted version of the preceding row, i.e., 
\[ 
A = \begin{bmatrix}a_0 & a_1 & \cdots & a_2 & a_1 \cr a_1 & a_0  & a_1 & & a_2\cr \vdots & a_1 & a_0 & \ddots & \vdots \cr a_2 & & \ddots & \ddots & a_1 \cr a_1 & a_2 & \ldots & a_1 & a_0  \end{bmatrix}.
\] 
The matrix $A$ is prescribed by selecting the values of $a_0,a_1,\ldots, a_{\lfloor n/2\rfloor}$ as either $0$ or $1$. We assume throughout that $a_0=0$ so that no oscillator is coupled to itself.

The eigenvalues of any circulant matrix (with $a_0=0$) are known analytically~\cite{golub2012matrix} and are given by 
\begin{equation} 
\lambda_p(A) = \sum_{s=1}^{n-1} a_s e^{2\pi i p s/n},  \quad 0\leq p\leq n-1, 
\label{eq:eigenvalues} 
\end{equation} 
where $a_{s} = a_{n-s}$ for $\lfloor n/2\rfloor<s< n$. The reflectional symmetry $a_{s} = a_{n-s}$ implies that the eigenvalues in~\eqref{eq:eigenvalues} are real and $\lambda_p(A) = \lambda_{n-p}(A)$ for $1\leq p\leq n-1$. Moreover, the eigenvectors of $A$ are given by~\cite{golub2012matrix}
\begin{equation} 
\begin{aligned}
A\underline{v}_p &= \lambda_p(A)\underline{v}_p, \quad 0\leq p \leq \lfloor n/2\rfloor, \\
A\underline{w}_p &= \lambda_p(A) \underline{w}_p, \quad 1\leq p \leq \lceil n/2\rceil-1, 
\label{eq:eigenvectors} 
\end{aligned} 
\end{equation} 
where 
\[
\underline{v}_p = \begin{bmatrix} 1 \cr \cos\!\left(\frac{2\pi p}{n}\right)\cr \vdots \cr \cos\!\left(\frac{2\pi p(n-1)}{n}\right)  \end{bmatrix}, \quad \underline{w}_p = \begin{bmatrix} 0 \cr \sin\!\left(\frac{2\pi p}{n}\right)\cr \vdots \cr \sin\!\left(\frac{2\pi p(n-1)}{n}\right)  \end{bmatrix}.
\]

Several equilibrium points of~\eqref{eq:dynamical} can be derived from linear algebra when $A$ is a circulant matrix. Let $\underline{\theta} = (\theta_0,\ldots,\theta_{n-1})^\top$, $\underline{c} =  \cos(\underline{\theta})^\top$, and $\underline{s} =  \sin(\underline{\theta})^\top$, where the superscript $\top$ denotes the vector transpose. Then since $\sin(x-y) = \sin(x) \cos(y) -\cos(x) \sin(y)$, a vector $\underline{\theta}$ is an equilibrium point of~\eqref{eq:dynamical} if and only if
\[
D_{\underline{c}} A \underline{s} =  D_{\underline{s}} A \underline{c}, \quad D_{\underline{c}} = {\rm diag}(\underline{c}), \quad D_{\underline{s}} = {\rm diag}(\underline{s}), 
\]
where ${\rm diag}(\underline{c})$ is a diagonal matrix with $\underline{c}$ on the diagonal. 
Due to the structure of~\eqref{eq:dynamical}, if $\underline{\theta}$ is an equilibrium point, then it remains an equilibrium point if all its arguments are shifted by the same angle $\Theta$. To avoid these trivial rotations, we set $\theta_0 = 0$.  Under this restriction, we find from~\eqref{eq:eigenvectors} that
\begin{equation} 
\underline{\theta}^{(p)} = \left(0,\frac{2\pi p}{n},\ldots,\frac{2\pi p(n-1)}{n}\right)^\top 
\label{eq:twisted} 
\end{equation} 
is an equilibrium point of~\eqref{eq:dynamical} for any $0\leq p\leq \lfloor n/2\rfloor$. When $p=0$, $\underline{\theta}^{(0)} =(0,\ldots,0)^\top$ and we call this the synchronous state, corresponding to all the oscillators being in phase. We call~\eqref{eq:twisted} a twisted state for $1\leq p\leq \lfloor n/2\rfloor$ as such a state physically corresponds to arranging the oscillators so that their phases differ by a constant amount from one oscillator to the next, twisting uniformly through $p$ full revolutions as we circulate once around the network.  For any circulant network, the twisted states are always equilibrium points but there are usually other equilibria too~\cite{canale2015exotic}.

To investigate the stability of the twisted states, we take a look at the Jacobian associated to~\eqref{eq:dynamical}. It is given by, for $0\leq p\leq \lfloor n/2\rfloor$,
\[
(J_p)_{jk} = \begin{cases} A_{jk} \cos\!\left(\theta_k^{(p)} - \theta_j^{(p)}\right), & j\neq k, \cr -\sum_{s=0}^{n-1} A_{js} \cos\!\left(\theta_s^{(p)} - \theta_j^{(p)}\right), & j = k.  \end{cases}
\]
It can be verified that $J_p$ is a symmetric circulant matrix for $0\leq p\leq \lfloor n/2\rfloor$.  
%\begin{itemize}[leftmargin=*,noitemsep]
%\item Symmetric: $J_p$ is symmetric because $A$ is a symmetric matrix and cosine is an even function.
%\item Circulant: First, ${\rm diag}(J_{p})={\rm diag}(AC)$, where $C_{jk} = \cos(\theta_k^{(p)} - \theta_j^{(p)})$. Since $A$ and $C$ are symmetric circulant matrices, their matrix product is too and hence the diagonal entries of $AC$ are constant. Second, the off-diagonal entries of $J_p$ are from a symmetric circulant matrix because $A$ and $C$ are symmetric circulant matrices and hence so is their Hadamard product.
%\end{itemize}  
This means that the eigenvalues of $J_p$ at the equilibrium points in~\eqref{eq:twisted} can be analytically derived. For any $0\leq p\leq \lfloor n/2\rfloor$, we have~\cite{golub2012matrix} 
%\begin{equation} 
%\lambda_r(J_p) = \sum_{s=1}^{n-1} a_s %\cos\!\left(\frac{2\pi p s}{n}\right)\!\left(-1 +  e^{2\pi i r s/n}\right)
%\label{eq:formula} 
%\end{equation} 
%for $0\leq r\leq n-1$. Since $J_p$ is a real symmetric matrix, the eigenvalues of $J_p$ are real and we can safely take the real part of the right-hand side of~\eqref{eq:formula}. We obtain
\begin{equation}
\lambda_r(J_p) = \sum_{s=1}^{n-1} a_s \cos\!\left(p t_s\right)\left[-1+\cos\!\left(r t_s\right)\right],
\label{eq:usefulFormula}
\end{equation} 
where $t_s = 2\pi s/n$.  The stability of the twisted state~\eqref{eq:twisted} depends on the signs of $\lambda_0(J_p),\ldots,\lambda_{n-1}(J_p)$. If one of these numbers is positive, then $\underline{\theta}^{(p)}$ is unstable. The matrix $J_p$ will always have at least one zero eigenvalue because of the trivial rotations by $\Theta$ mentioned earlier; if all other eigenvalues are negative, then $\underline{\theta}^{(p)}$ is stable. 
%Since $J_0$ is not the zero matrix and $\lambda_r(J_0)\leq 0$ for every $0\leq r\leq n-1$, we conclude that the synchronous state is always stable. 

To illustrate the general theory, we begin by applying it to two familiar examples of circulant networks.

%\paragraph{Complete graph.} 
%When $A$ is the adjacency matrix for a complete graph, then $a_1=\cdots =a_{\lfloor n/2\rfloor} = 1$. By the discrete orthogonality of cosines, we find that
%\[
%\lambda_r(J_0) = \begin{cases}-n, & r \neq 0, \\ 0, & r=0,\end{cases} \,\,\; %\lambda_r(J_p) = \begin{cases}0, & r\neq p, n-p,\cr n/2 & r = p, n-p. %\end{cases}
%\]
%The presence of the positive eigenvalue $n/2$ implies that all the twisted states are unstable. Hence the synchronous state is the only stable equilibrium point of the form~\eqref{eq:twisted}. In fact, it is a global attractor in this case~\cite{watanabe1994constants}. 

\subsection{Ring network} 
Consider a ring of oscillators with nearest-neighbor coupling so that $a_1 = 1$ and $a_k=0$ for $2\leq k\leq \lfloor n/2\rfloor$.  We find that
\[
\begin{aligned} 
\lambda_r(J_p) & = 2\cos\!\left( \frac{2\pi p}{n} \right)\!\left[-1+\cos\!\left( \frac{2\pi r}{n} \right)\!\right].\\
%&= -4\cos\left( \frac{2\pi p}{n} \right)\sin\left(\frac{\pi r}{n}\right)^2.
\end{aligned} 
\]
Hence $\underline{\theta}^{(p)}$ is unstable if and only if the twist $p$ satisfies $n/4<p\leq \lfloor n/2\rfloor$. 

\subsection{The Wiley--Strogatz--Girvan (WSG) network~\cite{wiley2006size}.} 
This is the network with adjacency matrix given by $a_1=\cdots = a_{\ell}=1$, where $1\leq \ell <\lfloor n/2\rfloor$. In other words, each oscillator on a ring is connected to its $\ell$ nearest neighbors on either side. Trigonometric identities yield  
\[
\begin{aligned} 
\lambda_r(J_p) &= 2\sum_{s=1}^{\ell}\cos\!\left( \frac{2\pi p s}{n} \right)\!\left[-1+\cos\!\left(\frac{2\pi r s}{n}\right)\right] \\
& = \frac{1}{2}U_{2\ell}(x_{p+r}) +  \frac{1}{2}U_{2\ell}(x_{p-r}) - U_{2\ell}(x_p),
\end{aligned} 
\]
where $U_{2\ell}$ is the degree $2\ell$ Chebyshev polynomial of the second kind (see Table~18.3.1 in Olver \textit{et al.}~\cite{olver2010nist}) and $x_p = \cos(p\pi/n)$. From this formula, we can numerically verify that there is a sequence of networks with connectivity tending to 68.09\% for which the 1-twist is a stable equilibrium. This matches the results derived in Wiley \textit{et al.}~\cite{wiley2006size}. Canale and Monz\'{o}n~\cite{canale2015exotic} formed a sequence of networks from the lexicographic product of a WSG network together with complete graphs to show that $\mu_c\geq 15/22\approx 0.6818$.

\section{\label{sec:Dense}Dense circulant networks that avoid (or almost avoid) global synchrony} 
\begin{figure} 
\begin{overpic}[width=.49\textwidth]{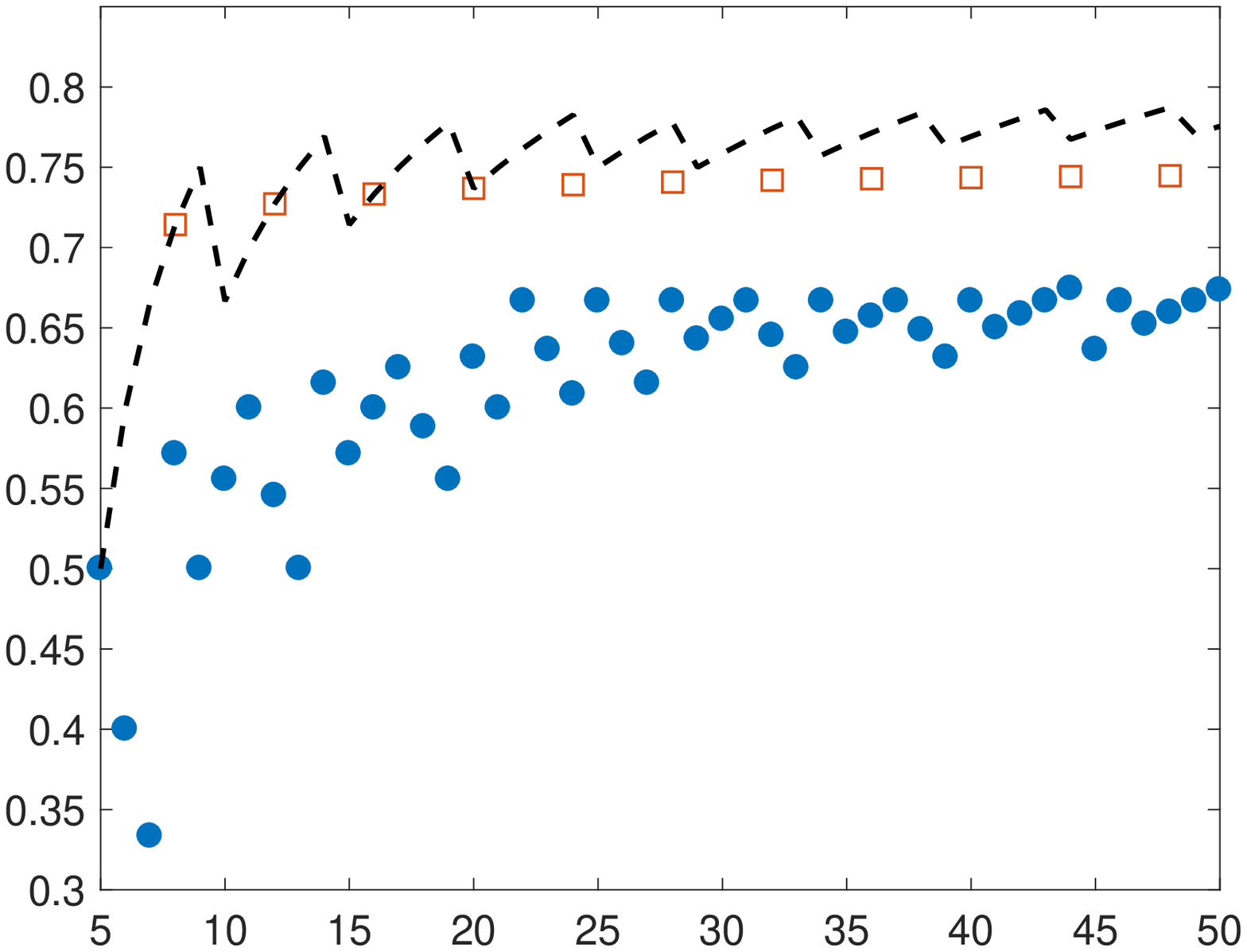}
\put(35,0) {$n = \text{size of network}$}
\put(0,27) {\rotatebox{90}{{$\delta_G/(n-1)$}}}
\put(17,64) {Bound from [Ling, Xu, and Bandeira, 2018]}
\end{overpic} 
\caption{Connectivity of the densest circulant networks with a twisted state that is either stable or weakly unstable for $5\leq n\leq 50$. The blue dots indicate networks with at least one stable twisted state. (All the nontrivial eigenvalues are negative.)  The red squares show a sequence of dense circulant networks with a twisted state that has all negative eigenvalues apart from 4 zero eigenvalues. The dashed line is a rigorous bound for each $n$, above which no stable states (of any kind) exist apart from synchrony; this conclusion follows from Theorem~3.1 of Ling \textit{et al.}~\cite{ling2018landscape}}
\label{fig:Densenetworks}
\end{figure}

Having reviewed circulant networks, we now use them to look for dense networks that have other attractors besides perfect synchrony, and which can therefore avoid having to globally synchronize. Every vertex in a circulant network $G$ has the same degree, $\delta_G$, so an equivalent aim is to maximize $\delta_G/(n-1)$ over all networks that fail to globally synchronize. To do this, we conducted a numerical search over all circulant networks of size $5\leq n\leq 50$ having at least one stable twisted state (for which all nontrivial eigenvalues are negative). The blue dots of Fig.~\ref{fig:Densenetworks} show the maximum value of $\delta_G/(n-1)$ for each $n$. They all lie below the best known lower bound~\cite{canale2015exotic} of $15/22\approx 0.6818$. But, notice the red squares that are far above the blue dots. These represent networks perched on the razor's edge of being able to avoid global synchrony. This sequence of remarkable networks squeezes up to 75\% connectivity as $n\rightarrow \infty$. Half of these networks have size $n = 8m$ and the other half have size $n = 8m+4$. (The stability properties for these two cases are similar so from now on we restrict our attention to $n = 8m$.)

\subsection{A sequence of networks on the razor's edge}
For $n = 8m$, consider the circulant network $G$ such that $a_j = 1$ if and only if ${\rm mod}(j,4)\neq 2$ for $1\leq j\leq 4m$. The degree of each vertex is $\delta_G = 6m-1$. The twisted state $\underline{\theta}^{(2m)}$ in~\eqref{eq:twisted} has no positive eigenvalues for this network. To see this, use~\eqref{eq:usefulFormula} and the identity $-1+\cos(2x) = -2\sin^2(x)$ to obtain 
%\[
%\begin{aligned} 
%\lambda_r(J_{2m}) & = -2\sum_{s=1}^{2m-1} \cos\!\left(\frac{2\pi (2m)(4s)}{8m}\right) \sin\!\left(\frac{\pi (4s) r}{8m}\right)^2 \\
%& -\cos\!\left(2\pi m\right) \sin\!\left(\pi r\right)^2\\
%& -2\sum_{s=0}^{2m-1} \!\! \cos\!\left(\frac{2\pi (2m)(4s+1)}{8m}\right) \!\sin\!\left(\frac{\pi (4s+1) r}{8m}\right)^2\\
%& -2\sum_{s=0}^{2m-1} \!\!\cos\!\left(\frac{\pi (2m)(4s+3)}{8m}\right)\!\sin\!\left(\frac{\pi (4s+3) r}{8m}\right)^2.
%\end{aligned} 
%\]
%Canceling the common terms, we find that
\[
\begin{aligned} 
\lambda_r(J_{2m}) &= -2\sum_{s=1}^{2m-1} \cos\!\left(2\pi s\right) \sin^2\!\left(\frac{\pi s r}{2m}\right) \\
%& -\cos\!\left(2\pi m\right) \sin\!\left(\pi r \right)^2\\
& -2\sum_{s=0}^{2m-1} \cos\!\left(2\pi s + \pi/2\right) \sin^2\!\left(\frac{\pi (4s+1) r}{8m}\right)\\
& -2\sum_{s=0}^{2m-1} \cos\!\left(2\pi s + 3\pi/2\right) \sin^2\!\left(\frac{\pi (4s+3) r}{8m}\right).
\end{aligned} 
\]
Noting that $\cos\!\left(2\pi s + \pi/2\right) = \cos\!\left(2\pi s + 3\pi/2\right) = 0$ and $\cos\!\left(2\pi s\right) = 1$ for any integer $s$, we conclude that
\[
\lambda_r(J_{2m}) = -2\sum_{s=1}^{2m-1} \sin^2\!\left(\frac{\pi s r}{2m}\right)  
%- \sin\!\left(\pi r \right)^2 
\leq 0, \quad 0\leq r\leq n-1.
\]
This means that $J_{2m}$ has all negative eigenvalues apart from four that vanish (corresponding to $r = 0, 2m, 4m, 6m$), as promised earlier. We have found a dense network of size $8m$ where the stability or instability of the $2m$-twist cannot be determined by the eigenvalues of the Jacobian.  Unfortunately, these twisted states turn out to be weakly (nonlinearly) unstable, and so the system does not manage to avoid global synchrony; otherwise, this sequence would have dramatically improved the lower bound on $\mu_c$ from $0.6818$ to $0.75$.

\begin{figure} 
\begin{overpic}[width=.49\textwidth]{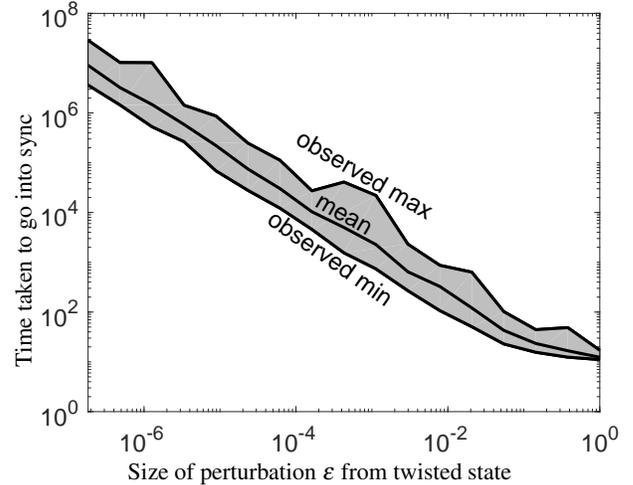}
\put(19,-2) {Size of perturbation $\epsilon$ from twisted state}
\put(2,15) {\rotatebox{90}{{Time taken to go into sync}}}
\end{overpic} 
\caption{Slow escape from a weakly unstable twisted state for $n = 8$. We integrated~\eqref{eq:dynamical} starting from random initial conditions within a ball of size $\epsilon$ about the twisted state~\eqref{eq:twisted} for $p = 2$. For each $\epsilon$, we repeated the simulation 100 times, to gather statistics, and computed the time taken to reach the in-phase state to within machine precision.}
\label{fig:LongTimeSimulations}
\end{figure} 

Figure~\ref{fig:LongTimeSimulations} shows the consequences of this weak instability. For initial conditions starting within an $\epsilon$-neighborhood of the $2m$-twisted state, it takes an algebraically long time of $\mathcal{O}(\epsilon^{-1})$ to escape that neighborhood before settling into the synchronous state. This super-slow escape time hints that these networks are poised at criticality. Perhaps, just a small tweak --- in the network topology, the sine coupling function, or the edge weights --- would be enough to nudge the system away from being globally synchronizing.  In any case, this leads us to conjecture that $\mu_c = 0.75$.

\subsection{An improved lower bound}
To sidestep the subtleties associated with the presence of zero eigenvalues, let us now restrict attention to circulant graphs with at least one certifiably stable twisted state; here, stable means that all its nontrivial eigenvalues lie strictly in the left half plane. For circulant graphs of size $5\leq n \leq 50$, the densest such graph is found at $n=44$ with $\delta_G/(n-1) \approx 0.6744$. This graph does not globally synchronize, of course, since it has a competing attractor (the stable twisted state), and it allows us to construct a sequence of even denser graphs with the same property, as follows. Given any non-globally-synchronizing graph $G$ and the complete graph $K_\tau$ (for any integer $\tau$), Canale and Monz{\'o}n~\cite{canale2015exotic} proved that the lexicographic product $G[K_\tau]$ is also non-globally-synchronizing.  Intuitively, the lexicographic product of $G$ and $K_\tau$ is formed by replacing each node in $G_1$ by a clique by size $\tau$; the nodes in different cliques are then connected in the same way as the parent nodes they replaced. We call this process twinning. To be more precise, the vertex set of the twinned graph $G[K_\tau]$ is defined as the Cartesian product $G\times K_\tau$, and any two vertices $(u,v)$ and $(x,y)$ are adjacent if and only if either $u$ is adjacent with $x$ in $G$ or $u = x$. The twinned graph $G'=G[K_\tau]$ is denser than $G$, because $\delta_G' = \tau-1+\tau \delta_G$ and $\delta_G'/(\tau n -1) > \delta_G/(n-1)$ for $\tau>1$. Thus, given any non-globally-synchronizing graph $G$, one can use this idea to construct a sequence of non-globally-synchronizing graphs with a limiting edge density of $(\delta_G +1)/n$. Canale and Monz{\'o}n~\cite{canale2015exotic} used this to show that the critical connectivity $\mu_c$ is bounded below by $15/22\approx 0.6818$. %Of course, one can start from any non-self-syncing graph (small and dense are best). 

Unfortunately, we have not been able to improve on the lower bound with just twinning. However, with the help of a computer search, we have found a more elaborate $4$-step ``twinning plus adding'' construction that slightly improves the lower bound to $0.68284$:  
(i) Take the $n=22$ WSG network with degree $14$, where the $1$-twist is observed to be stable, based on its eigenvalues. We will denote this graph by $G_{22}$ and label its vertices $v_1,\ldots,v_{22}$ so that $v_1$ is connected to $v_2,\ldots,v_8$. (ii) Now, form the graph $G_{22}[K_{85}]$ by twinning. This is a non-globally-synchronizing graph of size $n = 1870$ with degree $1274$. This graph is just like $G_{22}$ except vertex $v_k$ is now a connected cluster $C_k$ of size $85$. (iii) Pair up vertices from $C_1$ with vertices from $C_{9}$ (so that every vertex is paired up precisely once). Repeat this process with $C_2$ and $C_{10}$, $C_{3}$ and $C_{11}$, and so on, up to $C_{22}$ and $C_{8}$. Each vertex gets two additional edges, and the graph is still circulant. We now have a circulant graph $G'$ of size $n = 1870$ with degree $1276$. By looking at the eigenvalues of the associated Jacobian, the graph is observed to remain non-globally-synchronizing (with the same stable equilibrium state from $G_{22}[K_{85}]$). (iv) One can finally consider the sequence $G'[K_1],G'[K_2],\ldots$ of non-globally-synchronizing twinned graphs. The connectivity of this sequence of graphs tends to $1277/1870 \approx 0.6828$. (A computer search shows that $n = 22$ and $\tau = 85$ are the best parameters in the range of $5\leq n\leq 50$ for this construction.)  

%\paragraph{Small non-self-synchronizing graphs.}
%
%\begin{table} 
%\tabular{ccc}
%n & Degree & Graph \\ 
%4 & 2 & \\
%7 &  &  \\
%8 &  & \\
%9 &  & \\
%10 &  & \\
%\end{tabular} 
%\caption{}
%\end{table} 
%\newpage
\section{\label{sec:Sparse}Sparse circulant networks that (probably) globally synchronize }

\begin{figure} 
\begin{overpic}[width=.49\textwidth]{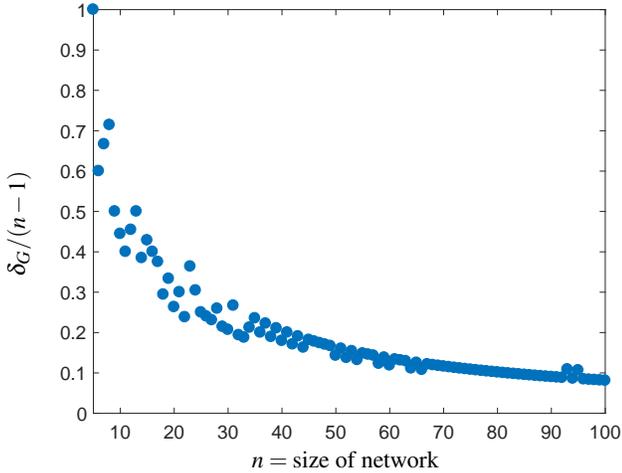}
\put(37,0) {$n = \text{size of network}$}
\put(0,29) {\rotatebox{90}{{$\delta_G/(n-1)$}}}
\end{overpic} 
\caption{Connectivity of the sparsest circulant networks for which all the twisted states are unstable. The trend observed here led us to a particular sequence of sparse circulant networks, all of which appear to be globally synchronizing.}
\label{fig:Sparsenetworks}
\end{figure}

As a counterpoint to the search for dense networks that fail to globally synchronize, we have also looked for sparse networks that do globally synchronize. Figure~\ref{fig:Sparsenetworks} shows, for $5\leq n\leq 100$, the sparsest circulant networks that lack any stable twisted state as revealed by a brute-force numerical search. The trend in Fig.~\ref{fig:Sparsenetworks} is $\delta_G = \mathcal{O}(\log_2 n)$.  Hence, the connectivity $\delta_G/(n-1)$ of these networks tends to zero as $n \rightarrow \infty$.

To understand this logarithmic trend, consider the circulant network of size $n=2^m$, where $a_j = 1$ for all powers of $2$ less than or equal to $2^{m-1}$ (Fig.~\ref{networkExamples}, bottom row). The other $a_j$'s are zero, or determined by the symmetry condition $a_{s} = a_{n-s}$ for $1\leq s\leq n/2$. This network is constructed by connecting a logarithmically small number of neighbors to each vertex of a ring. The final network has a total of $(2\log_2 n-1)n/2$ edges. 

All the twisted states $\underline{\theta}^{(1)}, \ldots, \underline{\theta}^{(n-1)}$ are unstable for this network. To prove this, we need to show that each Jacobian $J_1,\ldots, J_{n/2}$ has at least one strictly positive eigenvalue. Consider the $p$th eigenvalue of $J_p$. By~\eqref{eq:usefulFormula} and the fact that $\cos (x)[-1+\cos(x)] = [1-2\cos(x) +\cos(2x)]/2$, we find that 
\[
\begin{aligned} 
\lambda_p(J_p) &= m + \sum_{k=0}^{m-2} \cos\left(\frac{2\pi p 2^{k+1}}{2^m}\right) - 2 \sum_{k=0}^{m-2} \cos\left(\frac{2\pi p 2^k}{2^m}\right)\\&  - \cos(p\pi).
%\begin{aligned} 
%\lambda_p(J_p) &= m-1 + \sum_{k=0}^{m-2} \cos\left(\frac{2\pi p 2^k}{2^m}\right)
%- 2 \sum_{k=0}^{m-2} \cos\left(\frac{2\pi (2p) 2^k}{2^m}\right)\\& %+ \frac{1}{2} - \cos(p\pi) + \frac{1}{2}. \\
%& \qquad + \sum_{k=0}^{m-2} \cos\left(2p\frac{2\pi 2^k}{2^m}\right) + \frac{1}{2}\cos\left(2p\frac{2\pi 2^{m-1}}{2^m}\right).
%& = -2\cos\left(p\frac{2\pi}{2^m}\right) - \sum_{k=1}^{m-2} \cos\left(p\frac{2\pi 2^k}{2^m}\right) - \cos(p\pi) + m-\frac{1}{2} + \cos(p\pi) + \frac{1}{2}\\
\end{aligned} 
\]
This expression can be simplified to 
\begin{equation} 
\begin{aligned} 
\lambda_p(J_p) &= m -\sum_{k=1}^{m-2} \cos\!\left(\frac{2\pi p 2^k}{2^m}\right) - 2\cos\!\left(\frac{2\pi p}{2^m}\right) \\ %   -2\cos\left(p\frac{2\pi}{2^m}\right) - \sum_{k=1}^{m-2} \cos\left(p\frac{2\pi 2^k}{2^m}\right) + m\\
& \geq m- (m - 2) - 2  = 0,
\end{aligned} 
\label{eq:sparseFormula} 
\end{equation} 
where we can only hope to have an equality in~\eqref{eq:sparseFormula} if $\cos\left(2\pi p/{2^m}\right) = 1$, 
%if $\cos\!\left(\frac{2\pi p}{2^m}\right) = 1$, 
i.e., if $2^{-m}p$ is an integer. Since $0\leq p\leq 2^m-1$, we conclude that 
%$\cos\!\left(\frac{2\pi p}{2^m}\right)<1$ 
$\cos\left(2\pi p/{2^m}\right)<1$ 
for $1\leq p\leq 2^m-1$. In other words, $\lambda_p(J_p) >0$ for $1\leq p\leq 2^m-1$. Thus, all the twisted states are unstable. 

The network we have just constructed is very sparse.  We conjecture that if a circulant network is even more sparse, namely $\delta_G \ll \log_2 n$ as $n \rightarrow \infty$, then at least one of the twisted states must become stable. 
%(Any connected tree globally synchronizes so the circulant assumption is crucial here.) 
We go further and conjecture that our sparse circulant examples not only lack stable twisted states; they lack competing attractors of \emph{any} kind. Proving this conjecture for all $n = 2^m$ seems difficult because it requires checking that synchrony is the only stable state of~\eqref{eq:dynamical}. The conjecture holds for $n=2$ and $n=4$ because then the networks are complete graphs, which are known to be globally synchronizing~\cite{watanabe1994constants}. For $n = 8$, we used the computational algebraic geometry package {\it Macaulay2} \cite{M2} to confirm that there are $262$ real isolated equilibrium points of~\eqref{eq:dynamical} and all these equilibria are unstable except for the synchronous state. (There are an additional $1,\!008$ isolated complex-valued equilibria, and also 5 continuous families of unstable equilibria.) For larger $n$, the number of real equilibria grows exponentially. For $n = 16$, there are at least $32,\!768$ real equilibria, but there could be as many as $1,\!073,\!741,\!824$ complex-valued ones. The lesson is that a brute-force search quickly becomes computationally hopeless. A new idea is needed.

\section{\label{sec:Discussion}Discussion} 
%Despite the unfinished nature of the theory presented here (or perhaps because of it), we hope our work will stimulate further studies of these matters. 
Much remains to be learned about the conditions that ensure global synchrony, as well as the conditions that allow coupled oscillators to avoid it. Even in the simplified setting of  Kuramoto oscillators, many fascinating problems remain about the impact of network topology on global synchronization. We have highlighted two such problems in the conjectures mentioned above. Similar questions could be investigated for other kinds of oscillators, such as Kuramoto-Sakaguchi oscillators, van der Pol oscillators, Stuart-Landau oscillators, and so on.      

The study of global synchronization can also have practical implications. Existing theorems for a different class of oscillators (pulse-coupled oscillators)~\cite{mirollo1990synchronization} have been applied to sensor networks and distributed clock synchronization~\cite{simeone2008distributed,yick2008wireless,hong2005scalable,werner2005firefly}, and to the design and implementation of ultra-low-power radio systems~\cite{dokania2011low} with potential uses that include intrusion detection, keeping track of workers in coal mines, personal health care, and automated drug-delivery~\cite{yick2008wireless}. In the same vein, the results presented here for Kuramoto oscillators may open up additional ways to build useful networks that reliably keep themselves in sync.

\begin{acknowledgments}
We thank Heather Harrington for discussions of the algebraic geometry computations; Mateo Diaz for informing us of~\cite{ling2018landscape}; Eduardo Canale for pointing out the weak instability of the critical twisted state in our dense circulant networks; and Lee DeVille, Gokul Nair, and Heather Wilber for helpful discussions.  Research supported by NSF Grant No.~DMS-1818757 to A.~T., NSF Grant No.~DMS-1502294 to M.~S., and NSF Grants No.~DMS-1513179 and No.~CCF-1522054 to S.~H.~S.
\end{acknowledgments}

\section*{Data Availability}
The data that supports the findings of this study are almost all available within the article. Any data that is not available can be found from the corresponding author upon reasonable request.

%\bibliography{apssamp}% Produces the biblionetworky via BibTeX.
%\bibliographystyle{unsrt}
%\providecommand{\noopsort}[1]{}\providecommand{\singleletter}[1]{#1}%

\end{document}